# First Principles Theories of Piezoelectric Materials

Ronald Cohen, Carnegie Institution of Washington, Washington, D.C. USA

cohen@gl.ciw.edu

Piezoelectrics have long been studied using parameterized models fit to experimental data, starting with the work of Devonshire in 1954. [1] Much has been learned using such approaches, but they can also miss major phenomena if the materials properties are not well understood, as is exemplified by the realization that low-symmetry monoclinic phases are common around morphotropic phase boundaries, which was missed completed by low-order Devonshire models, and can only appear in higher order models. [2] In the last 15 years, a new approach has developed using first-principles computations, based on fundamental physics, with no essential experimental input other than the desired chemistry (nuclear charges). First-principles theory laid the framework for a basic understanding of the origins of ferroelectric behavior [3-7] and piezoelectric properties. [8-11] The range of properties accessible to theory continues to expand as does the accuracy of the predictions. We are moving towards the ability to design materials of desired properties computationally. Here we review some of the fundamental developments of our understanding of piezoelectric material behavior and ability to predict a wide range of properties using theoretical methods. This is not meant as a review of the literature. Comprehensive reviews of the literature of theoretical studies of ferroelectrics are given by Resta [12] and Rabe and Ghosez. [13]

Most first-principles calculations for piezoelectrics are based on the density functional theory (DFT), [14] and most are within the local density approximation (LDA). [15] The energy for any configurations of atoms is computed by solving a set of effective Schrödinger equations with an effective potential that includes many-body contributions like those of a uniform electron gas at each point in space. Forces, phonon frequencies (via the dynamical matrix), effective charges, dielectric constants, elastic constants, piezoelectric constants, and polarization are all directly computable for the static lattice (zero temperature) for ordered structures.

To obtain properties for finite temperatures, it is necessary to use the primary first-principles results to parameterize an effective Hamiltonian or potential model, which can then be used study the effects of temperature and simulate disordered materials.

A number of major advances in our understanding of piezoelectric materials are due to first-principles studies. The first is the importance of hybridization. First-principles studies showed that ferroelectricity is due to the competition between long-range forces, which favor off-centering, and short-range forces that favor the high symmetry centric phase. [5] Unlike the venerable Slater rattling ion model, [16] the key element in oxide ferroelectrics is covalency, or hybridization between the cation and its oxygen neighbors that allows the cation to move off-center. This concept is now widely used in experiments and development of new piezoelectric materials. Secondly is the concept of polarization rotation, which is responsible for the giant electromechanical coupling seen in relaxor ferroelectrics such as $Pb(Mg_{1/3},Nb_{2/3})O_6$ (PMN)-$PbTiO_3$ (PT). [17] Thirdly is the relationship between cation ordering and polar nanoregions in relaxors. [18, 19] Fourth is the prediction of a morphotropic phase boundary in pure PT at high pressures with huge electromechanical coupling in the transition region, indicating that the main effect of the relaxor PMN or PZN, for example, is to tune the transition to zero pressure, rather than something intrinsic to relaxor behavior. [11] Fifth is the discovery of reentrant ferroelectricity, with ferroelectricity reappearing at very high pressures, indicating the possibility of whole new classes of ferroelectric materials. [20]

First-principles studies of ferroelectrics have also given rise to major advances in theoretical methods. The development of the modern theory of polarization [21-23] was motivated entirely to understand piezoelectrics, and has other broader implications as well. [24, 25] The theory of insulators under applied electric fields was also developed in order to understand piezoelectric materials. [26, 27]

The goal of computational research on piezoelectrics is three-fold: to help understand experiments, to help guide experiments, and to make predictions for new materials. An example of piezoelectric materials by design is exemplified by the predictions of interesting and tunable properties of ferroelectric superlattices. [28-31]

# I. The Origin of Piezoelectricity and Ferroelectricity: What we have learned from first-principles studies

Any insulating crystal with two or more sublattices that has the appropriate symmetry, i.e. is a member of one of the 20 polar crystal classes, will have a macroscopic polarization and be piezoelectric, but in order to have a large piezoelectric effect the crystal should contain ions with large effective charges and they should easily move as a result of lattice strains. All ferroelectrics are piezoelectric, but in addition must have a spontaneous polarization that is switchable by an applied electric field, and generally have a phase transition with increased temperature to a nonpolar, paraelectric state. Many perovskites are ferroelectric, and this review will consider mainly the oxide perovskites. Ferroelectricity in the $LiNbO_3$ structure is very similar in origin, but the structure allows uniaxial ferroelectricity only (the soft-mode is non-degenerate and the polarization must be along the c-axis), which significantly changes the dynamical behavior.

Ferroelectrics research was a very hot topic in the 1960's and earlier, driven by experimental studies and phenomenological models. Starting in the 1990's the field was rejuvenated with the introduction of first-principles methods along with a new generation of fundamental experimental studies.

## *A. First-principles methods*

First-principles methods are so called because they require no experimental data as input, and materials properties are computed using fundamental physics starting with electrons and nuclei and the Coulomb interactions among them. The exact solution of the Schrödinger equation (or its relativistic version the Dirac equation) is impossible for many electron systems because it is a 3N-dimensional differential equation, where N is of the order of $10^{23}$ for a bulk material. Almost all first-principles studies of ferroelectrics (and other materials as well) have used the Density Functional Theory (DFT), which reduces the problem to solution of a 3-d differential equation to find the charge density and total energy and its derivatives. The derivatives are used to compute elastic and piezoelectric constants, as well as vibrational frequencies. In principle, DFT is exact, but the exact functional is not known explicitly, so approximations are used. A typical and commonly used approximation is the local density approximation (LDA). In the LDA, the complicated many-body interactions among the electrons at each point in space in a material are modeled as being the same as the interactions in an electron gas of the same density as that density at that point. This is a surprisingly good approximation. Many tens of thousands of studies have been performed using these methods since the 1960s, and we now have a very good idea of their accuracy and reliability. Volumes are typically within a few percent of experiment, phase stability and phase transitions can be estimated reasonable, and many physical properties can be computed with reasonable accuracy, in some cases rivaling that of experimental studies. The small error in volume is not greatly significant in many materials, but for ferroelectrics it is cru-cial. This is because ferroelectrics are extremely sensitive to pressure (volume). Quite modest pressures destroy ferroelectricity in BaTiO$_3$ and PbTiO$_3$ corresponding to volume compressions of 2%, for example, so a small error in volume is very significant. As a result, most first-principles studies of ferroelectrics have used the experimental volume, or include a small shift in pressure of -1 to -5 GPa. New density functionals may remove this need to rely on experimental data. [32]

An important advance in our theoretical understanding of polar solids came through the development of the first-principles theory of polarization. [21-23] Contrary to what one finds in many textbooks, the polarization is not the dipole moment in a unit cell of a material, which in fact depends on the choice of cell. The polarization is rather computable as a Berry's phase from the wavefunctions, and can be thought of as a current flow from one structure to another. Thus it is the change in polarization that is computable. If the reference state is centrosymmetric, the polarization is the charge flow from that state to the final state of interest.

Another important advance was the development and implementation of the density functional perturbation or linear response theory, which allows direct computation of phonon frequencies as well as transverse effective charges, [33, 34] and very recently elastic and piezoelectric constants. [35] Such linear response methods effectively solve directly for derivatives of the Kohn-Sham equations to obtain derivatives of the energy with respect to perturbations in the potential. The dynamical matrix can be obtained directly in this way and diagonalized, giving the quasiharmonic phonon frequencies, which give insight into underlying crystal instabilities, can be compared with vibrational spectroscopy from Raman or Infrared spectroscopy experiments, and to inelastic neutron and X-ray scattering. A huge amount of information is contained in the phonon dispersion, particularly when coupled with the phonon eigenvectors, which are not generally available experimentally.

Linear response also allows the direct computation of the Born transverse effective charges, $Z^*$.[36] The effective charges are important for understanding piezoelectric behavior in a material, as well as for quantitative computations of piezoelectric constants. The effective charges give another clear indication of the importance of hybridization or atomic polarization. The transverse effective charges are related to the amount of charge that moves when a nucleus is displaced; it is the change in polarization with displacement of a nucleus, V d**P**/dr$_i$. The effective charge is a tensor, because the amount of charge that moves depends on the direction of the displacement. The transverse effective charges are also obtainable from optical experiments from the oscillator strengths or LO-TO-splitting. [37, 38]

Many important problems in piezoelectrics involve temperature or complex solid solutions, and cannot be treated efficiently using only first-principles methods directly. Effective Hamiltonians and potential models, both fitted to first-principles computations, allow the computation of anharmonic dynamical properties and non-equilibrium phenomena such as ferroelectric switching, as well as thermal properties for complex chemically disordered or heterogeneous materials. First-principles based effective Hamiltonians have been exploited especially well for ferroelectrics, starting with the pioneering work of Rabe and Joannopoulos. [7] In effective Hamiltonian models, only the lowest energy modes are considered, and a Hamiltonian is written in terms of the normal mode coordinate, the strain, and the strain-mode coupling. First-principles computations are then used to obtain the parameters of the Hamiltonian, and energy minimization, Monte Carlo, and/or molecular dynamics can then be used to obtain materials properties.[39-41]

## B. Applications to ferroelectrics

### 1. Perovskite ferroelectrics

First-principles calculations of the soft mode potential surface in perovskite $BaTiO_3$ and $PbTiO_3$ in 1989 showed that the LDA can predict ferroelectric behavior (Fig. 1). [4, 5] The total energy is a maximum for the perfect perovskite structure. For a cubic lattice, the energy is a minimum along [111] but tetragonal strain makes the minimum along [001] (Fig. 2). The strain coupling is critical for the tetragonal phases, and of course is primary for the piezoelectric effect. Strain has little effect in the rhombohedral phase under zero field, and to a good approximation the lattice remains cubic through the transition to a rhombohedra phase.

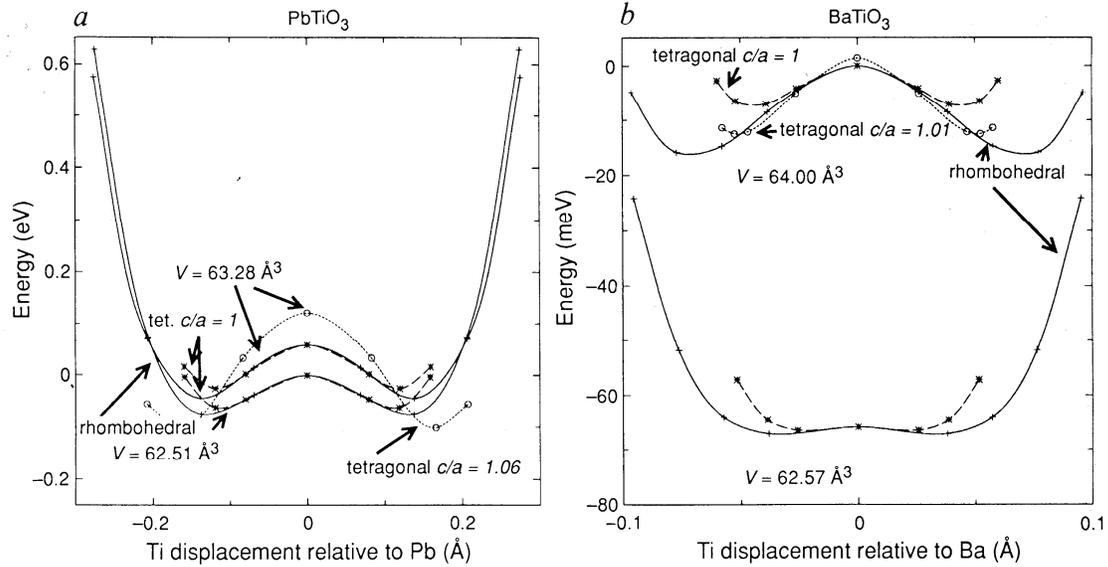

**Figure 1. Energy versus phonon displacement for (a) $PbTiO_3$ and (b) $BaTiO_3$ showing the multiple well potential surfaces that underlies ferroelectric behavior. Also illustrated is the great sensitivity to volume and shear strain. From Ref. [5].**

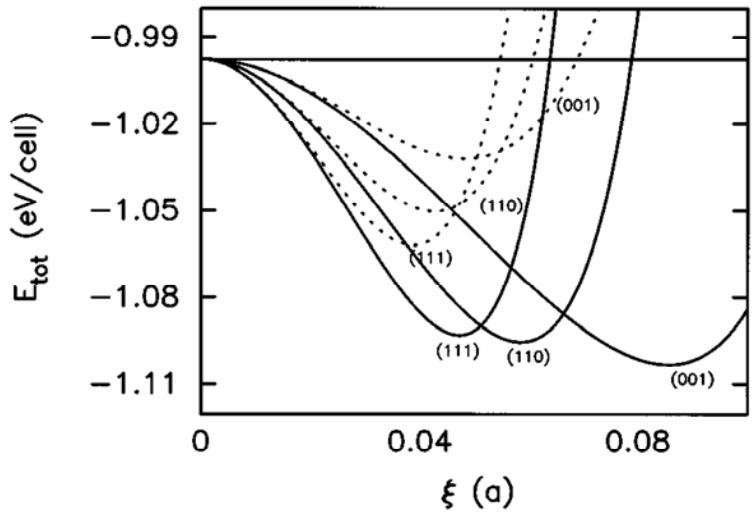

**Figure 2.** Energy versus more coordinate for PbTiO$_3$ for the unstrained, cubic lattice (dotted) and relaxed tetragonal lattice (solid). Note that the energy is a minimum along the (111), rhombohedral, direction for the unstrained lattice and along the (001), tetragonal, direction for the strained lattice. The great energetic importance of the lattice strain is also evident. This behavior contrasts with that of BaTiO$_3$, which would show (111) always most stable, which is why BaTiO$_3$ has a series of three phase transitions at zero pressures, whereas PbTiO$_3$ has only 1. From Ref. [41].

One of the long-standing questions was the origin of the ferroelectric effect. Slater presented a model for ferroelectricity in perovskite known as the "rattling ion model." [16] The idea is simply that the B-cation, e.g. Ti in BaTiO$_3$, is too small for the octahedral cage, and thus off-centers due to the electrostatic forces from the other ions, the local Lorentz force. The problem with this classic model is that it does not explain why almost all ferroelectric perovskites have B-cations which have d$_0$ electronic configuration, i.e. their lowest unoccupied conduction band orbital has d character, and no d-states are nominally filled. Examples of d$_0$ cations in B-sites in ferroelectrics are Ti$^{4+}$, Zr$^{4+}$, Nb$^{5+}$, Ta$^{5+}$, and Sc$^{3+}$.

First-principles calculations show clearly the following picture of the ferroelectric instability in oxide perovskites. Long-range Coulomb (Madelung) forces favor off-centering; there is a maximum in the ionic electrostatic energy for the atoms in the ideal perovskite positions. Short-range repulsive forces favor the ideal structure, where the atoms are as far apart as possible. Hybridization between the B cation d-states and oxygen p-states reduces the repulsion and allows the atoms to move off-center. The Pb "lone-pair" 6s states also lead to intrinsic off-centering of the Pb ions. The same picture holds in uniaxial ferroelectrics such as LiNbO$_3$ and LiTaO$_3$, [42] and may operate in many other ferroelectrics. Theoretical evidence for the above picture comes from studies charge densities and electronic densities of states for the ferroelectric and centrosymmetric structures. A clear indication is also the fact that if one removes the d-state degrees of freedom from the basis, the ferroelectric state is no longer stable.[4, 5]

Another matter of long-standing interest is the nature of the ferroelectric phase transition, whether it should be considered primarily as order/disorder or as a displacive phase transition. In the classical displacive picture, in the high temperature phase the high symmetry structure is considered as stable. As the phase transition temperature is approached from above the soft-mode frequency decreases. In a first-order transition the polarization, which is the primary order parameter suddenly acquires a non-zero value at the transition temperature, and the atoms move off-center. Simultaneously the soft mode hardens and increases in frequency with further reduction in temperature. In a second- or higher-order transition the order parameter varies continuously from zero to a non-zero value at the critical temperature $T_c$. Ferroelectrics tend to be weakly first-order. In a pure order-disorder transition, the off-center displacements persist above $T_c$, the main distinction being whether the displacements are ordered or disordered. Whether a transition will be displacive or order-disorder depends on the well depths relative to the temperature, and the coupling strength between displacements in one cell and its neighbors.

Experiments have long suggested that ferroelectric phase transitions are not simple displacive transitions. For example, the soft mode in BaTiO$_3$ or KNbO$_3$ in the cubic high temperature phase is triply degenerate, and at the transition to tetragonal one mode hardens, while the other two degenerate modes continue to soften as if nothing had happened. At the orthorhombic phase boundary one of these modes hardens and the other continues to soften. Only at the orthorhombic to rhombohedral boundary do all of the modes harden below the transition (Fig. 3). [43] Further experimental evidence that the cubic phase is not "simply cubic" comes from X-ray diffraction; instead of the simple Bragg diffraction spots for the cubic perovskite structure, a complex streaky pattern is observed. [44] Furthermore, EXAFS shows that bond distances do not vary across $T_c$, as would be expected in a displacive transition. [45] On the other hand, it has been known that the transition is not a pure order-disorder transition, because the entropy of transition is much reduced from what it would for a pure order-disorder transition. [46]

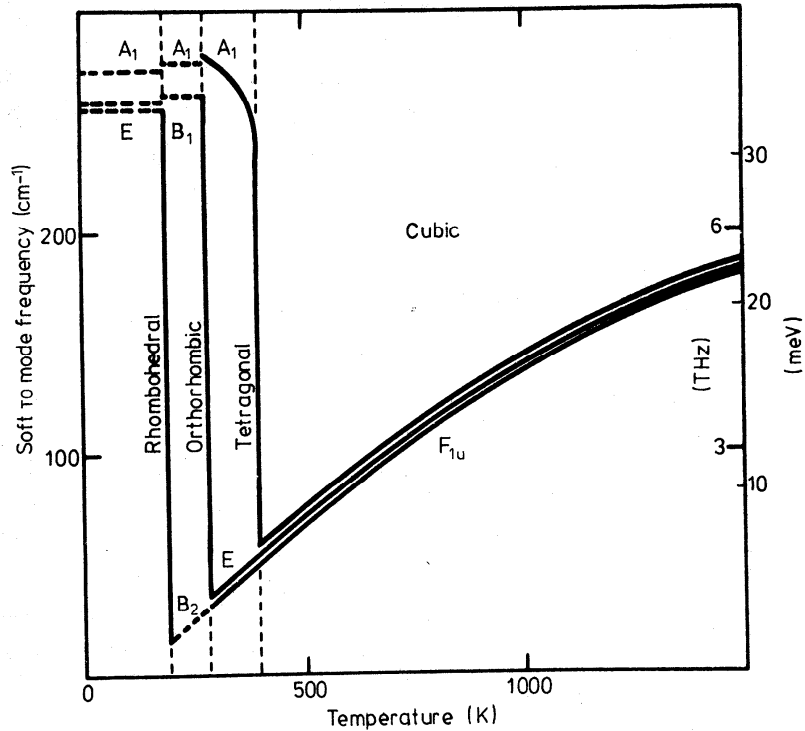

**Figure 3. Schematic of soft mode behavior in BatiO$_3$. One mode hardens at each transition, and the mode frequencies do not go to zero. Only the orthorhombic to rhombohedral transition looks similar to a conventional soft-mode, though even it should be understand as having a large order-disorder character. From Ref. [43].**

The first-principles static energy calculations from the beginning showed very deep wells along the [111] directions, the eight cube diagonals, giving rise to the eight-site model, and the order-disorder character was emphasized as had been proposed by Comes. [44] In the high temperature phase all eight sites are occupied, four are occupied in the tetragonal phase, two in orthorhombic, and one in rhombohedral, which is an ordered structure.

In the pure displacive model the soft mode is the lowest excitation related to the phase transition. In the pure order-disorder model, there is no soft mode, and the dynamics are dominated by hopping, giving rise to low frequency relaxational response. Dynamical simulations and experiments show that the true nature of the ferroelectric phases and phase transitions are intermediate between order-disorder and displacive. There is a soft mode and there is also a low frequency relaxational response. In different materials the relative intensity and importance of the relaxational and soft-mode response varies, and it varies in a single material as a function of temperature and pressure.

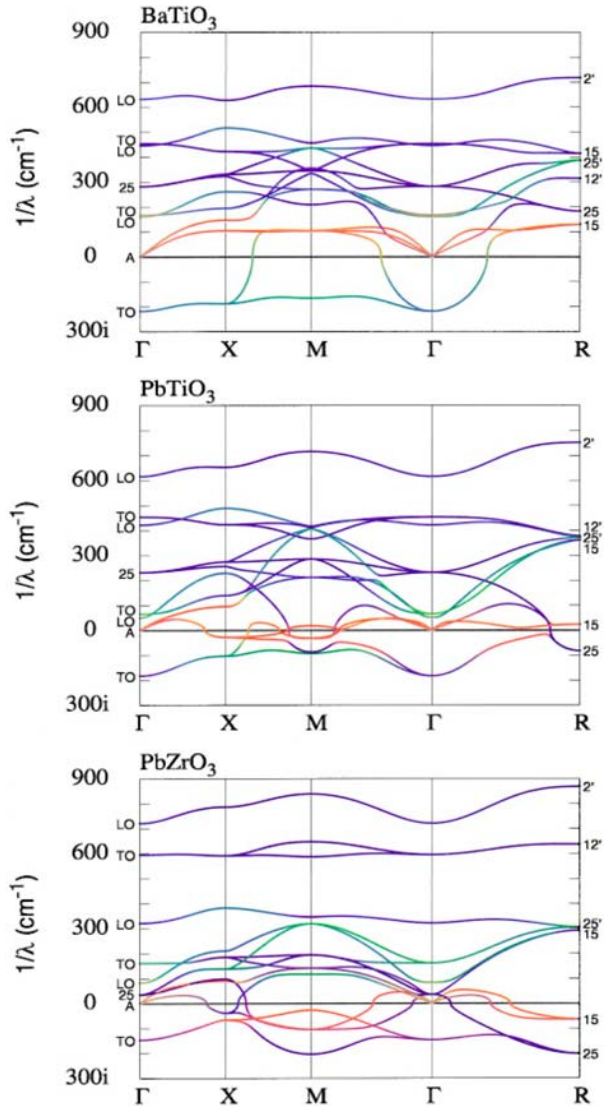

**Figure 4. Computed phonon dispersion curves for cubic BaTiO$_3$, PbTiO$_3$, and PbZrO$_3$. From Ref. [47].**

Linear response lattice dynamics calculations for the cubic structure show the underlying instabilities, and how they differ among ferroelectrics. Figure 4 shows the computed dispersion curves for the cubic structure of BaTiO$_3$, PbTiO$_3$, and PbZrO$_3$. [47] This type of information is completely unobtainable experimentally. Experiments give renormalized phonon frequencies (the actual dressed modes) but first-principles theory can be used to obtain the harmonic potential surfaces which illustrate the underlying instabilities in the structure. The frequencies on these plots actually represent the square root of the curvature of the potential surface as a function of wavevector. Unstable (imaginary) frequencies are plotted as negative numbers. Note the ferroelectric instability forms a flat band over most of the Brillouin zone for BaTiO$_3$ and PbTiO$_3$, but in the former it stiffens towards the R point, indicating strong stability against rotations and octahedral distortions in BaTiO$_3$. Both BaTiO$_3$ and PbTiO$_3$ have the most unstable mode at Γ, indicating the ferroelectric distortions are lower energy than antiferroelectric ones. PbZrO$_3$ on the other hand shows the most unstable points at R and M, i.e. rotational instabilities, consistent with its antiferroelectric ground state structure with octahedral tilts. The interplay of ferroelectric and rotational distortions has been studied in detail in PbZrO$_3$ and PZT. [48-51]

One of the great advances in understanding ferroelectrics was the development of the first-principles theory of polarization [21-23, 52]. Previously it was thought that polarization could be thought of as the dipole moment per unit cell divided by the unit cell volume. However, this quantity depends on the choice of unit cell. On second thought, many considered polarization to be determined as a difference between the dipole moment of the unit cell in a polar phase and the dipole moment of the unit cell in a centrosymmetric phase. This also doesn't work in general; one would have to know how the charge flows as the atoms displace from the non-polar to polar positions. In the case of a localized model, such as a shell model or localized Wannier functions, one still obtains a lattice of polarizations rather than a single value polarization. The polarization lattice has a spacing of e**R**/V, where **R** are the lattice vectors. These formal polarizations are not the quantities that are generally measured in experiments, which are always polarization differences; generally a current is measured in response to switching the polarization, for example. If one switches the polarization by 180$^o$ in a ferroelectric, the measured current flow corresponds to 2P. It is particularly counterintuitive that a centrosymmetric crystal can have a polarization lattice of (±**n**+½) e**R**/V where **n** are integers.

The reason for this behavior is that in periodic boundary condition, polarization is a Berry phase of the wave function; it depends on the *phase* of the wave function as a function of wavevector **k**, rather than the charge density, which depends on the square of the wave function. The polarization computed using the Berry's phase approach is the electronic part of the polarization, to which the atomic ion core contributions must be added. The claim that a localized model gives an unambiguous polarization [53] is not correct; even the completely localized Clausius-Mossotti model produces a lattice of compatible polarization in periodic boundary conditions.

Vanderbilt and King-Smith [22] showed that the polarization is related to the surface charge, and the different compatible formal polarizations correspond to different cell choices that can be consistent with different surface charges (modulo e**S**/A, where **S** is a surface lattice vector and A is the surface cell area. The centrosymmetric cases that have no **P** compatible with P=0 are where one cannot make a stoichiometric crystalline slab with the unit cell composition that has no dipole. For example, consider cubic BaTiO$_3$ with nominal charges, Ba$^{2+}$, Ti$^{4+}$, and O$^{2-}$. One can make a slab with BaO and TiO$_2$ surfaces with composition BaTiO$_3$ that has no dipole, and thus the lattice of polarizations includes P=0. However, cubic KNbO$_3$ with nominal charges, K$^+$, Nb$^{5+}$, and O$^{2-}$ gives one (KO)$^{1-}$ surface and one (NbO$_2$)$^{1+}$ surface, thus the polarization lattice is symmetric about 0, but the smallest values are (±½) e**R**/V. Real KNbO$_3$ would have defects or impurities to pacify the huge naked dipole across a perfect slab. Whether this difference in behavior is reflected in different surface, film, or bulk properties of ferroelectrics with paraelectric phases with formal polarization of (±**n**+½) e**R**/V versus those with formal polarization (±**n**) e**R**/V is an interesting question.

Usually the derivatives of the polarization are of most interest. The Born transverse effective charges, defined as:

$$Z^*_{s,\alpha\beta} = \frac{V}{e}\frac{\partial P_\alpha}{\partial u_{s,\beta}}$$

are greatly enhanced in oxide ferroelectrics, and the large coupling to electric fields is due partly to the softness of the potential surface, and partly due to the enhanced effectives charges, since the field **E** couples with displacements **d** via Sum **Z\* d**.

In ferroelectrics, the effective charges are often much larger than their nominal values, contrary to chemical intuition that they would be lower than their nominal charges due to cova-lency, as would be static Mulliken charges, for example. Tables 1 and 2 show computed effective charges for PbTiO$_3$ [8] and rhombohedrally ordered PMN [54]. The effective charges can be greatly enhanced, so that Z$^*$ for oxygen can reach values of above X! The fact that these enhanced charges come from hybridization was shown in the same way that the importance of hybridization was shown for the energies. When the d-states are removed from the basis, the effective charges regain nominal values. [55]

**Table 1. Computed effective charges for PbTiO$_3$. Note they are greatly enhanced due to hybridization over the nominal charges of +2, +4, and -2 for Pb, Ti, and O. From Ref. [8].**

| Atom | $Z_{xx}$ | $Z_{yy}$ | $Z_{zz}$ |
|---|---|---|---|
| Pb | 3.7 | 3.7 | 3.5 |
| Ti | 6.2 | 6.2 | 5.2 |
| O$_{(0,0.5,0.5)}$ | -2.6 | -5.2 | -2.2 |
| O$_{(0.5,0..5,0)}$ | -2.2 | -2.2 | -4.4 |

**Table 2. Computed effective charges for 1:2 ordered PMN with symmetry C2. From Ref. [54]. WP is the Wycoff position.**

| Atom Type | WP | $Z^*_{xx}$ | $Z^*_{yy}$ | $Z^*_{zz}$ | $Z^*_{xy}$ | $Z^*_{xz}$ | $Z^*_{yx}$ | $Z^*_{yz}$ | $Z^*_{zx}$ | $Z^*_{zy}$ |
|---|---|---|---|---|---|---|---|---|---|---|
| Pb | 1a | 3.34 | 3.93 | 3.35 | -0.29 | -0.24 | -0.50 | -0.50 | -0.23 | -0.28 |
| Pb | 2c | 3.57 | 3.79 | 3.50 | -0.13 | 0.00 | 0.09 | 0.58 | 0.24 | 0.53 |
| Mg | 1b | 2.89 | 2.87 | 2.88 | 0.04 | 0.14 | 0.02 | 0.02 | 0.14 | 0.04 |
| Nb | 2c | 6.39 | 6.41 | 5.70 | 0.03 | -0.39 | 0.11 | -0.26 | -0.13 | -0.08 |
| O1 | 2c | -3.39 | -2.74 | -2.32 | -0.11 | 0.43 | 0.05 | 0.12 | 0.43 | 0.03 |
| O2 | 2c | -1.95 | -2.19 | -4.03 | -0.06 | -0.19 | -0.04 | 0.03 | -0.19 | -0.10 |
| O3 | 2c | -2.39 | -3.69 | -2.34 | -0.20 | 0.16 | -0.35 | 0.19 | 0.15 | 0.33 |
| O4 | 2c | -2.08 | -2.27 | -4.79 | -0.03 | -0.56 | -0.10 | -0.19 | -0.14 | 0.12 |
| O5 | 1a | -2.09 | -5.37 | -2.07 | -0.13 | 0.31 | 0.20 | 0.19 | 0.31 | -0.15 |

The enhanced effective charges can also be considered to arise from atomic polarization comes from the shell model. The shell model is a mechanical model for materials, where atoms are modeled as charged "cores" and charged "shells" connected by springs. Atomic polarization is modeled by displacement of an atom's shell from its core. Shell models can reproduce large effective charge through the "knock-on" effect. For example, when an oxygen core is displaced, its shell is displaced. The shell interacts with other ionic shells through its short-range interactions (springs) and charge, and the other shells move as well. This can give a much greater change in polarization of the crystal than would occur in an isolated ion, and thus an enhanced effective charge. Effective charges can also be determined experimentally, from LO-TO splitting and from oscillator strengths. In fact, the enhancement in effective charges over their nominal charges was observed first experimentally, [37, 38] although its significance was not widely recognized.

Piezoelectric constants are derivatives of the polarization with respect to strain:

$$e_{\alpha\beta\delta} = \frac{\partial P_\alpha}{\partial \varepsilon_{\beta\delta}} = \sum_{s\mu} Z^*_{s,\alpha\mu} \frac{du_{s\mu}}{d\varepsilon_{\beta\delta}}.$$

The piezoelectric constants can be computed using a frozen phonon method, either by computing the **P** using the Berry phase method as a function of strain, or both first computing the effective charges and then relaxing the atomic positions as functions of strain to find the strain derivatives du/dε. [8-10] Wu and Cohen applied the new density functional perturbation theory [34] for elas-ticity and piezoelectricity [35] to PbTiO$_3$ and obtained the elastic and piezoelectric constants as functions of pressure (Table 3). [11] We found a set of transitions under pressure from tetragonal, to monoclinic, and then rhombohedral, like a morphotropic phase region seen in complex solid solutions like PZT (PbZrO$_3$-PbTiO$_3$) on in relaxor ferroelectrics. Furthermore, they we that the piezoelectric constants peaked in the transition region, giving electromechanical coupling larger than found in any known material (Figs. 5 and 6). This strongly suggests that large coupling seen in solid solutions with PbTiO$_3$ is simply due to chemical tuning of this transition to zero pressure rather than any complex structural or electronic effect.

**Table 3. Piezoelectric constants for PbTiO$_3$ at the experimental lattice computed using density functional perturbation theory (DFPT) and frozen strains (FS) [11] compared with experiment. From Ref. [11].**

| method | $E31$ | $e33$ | $e15$ | $c11$ | $c12$ | $c13$ | $c33$ | $c44$ | $c66$ |
|---|---|---|---|---|---|---|---|---|---|
| DFPT | 2.06 | 4.41 | 6.63 | 230 | 96.2 | 65.2 | 41.9 | 46.6 | 98.8 |
| FS | 2.07 | 4.48 | 6.66 | 229 | 95.6 | 64.3 | 41.2 | 47.2 | 98.6 |
| exp | 2.1 | 5.0 | 4.4 | 237 | 90 | 70 | 66 | 69 | 104 |

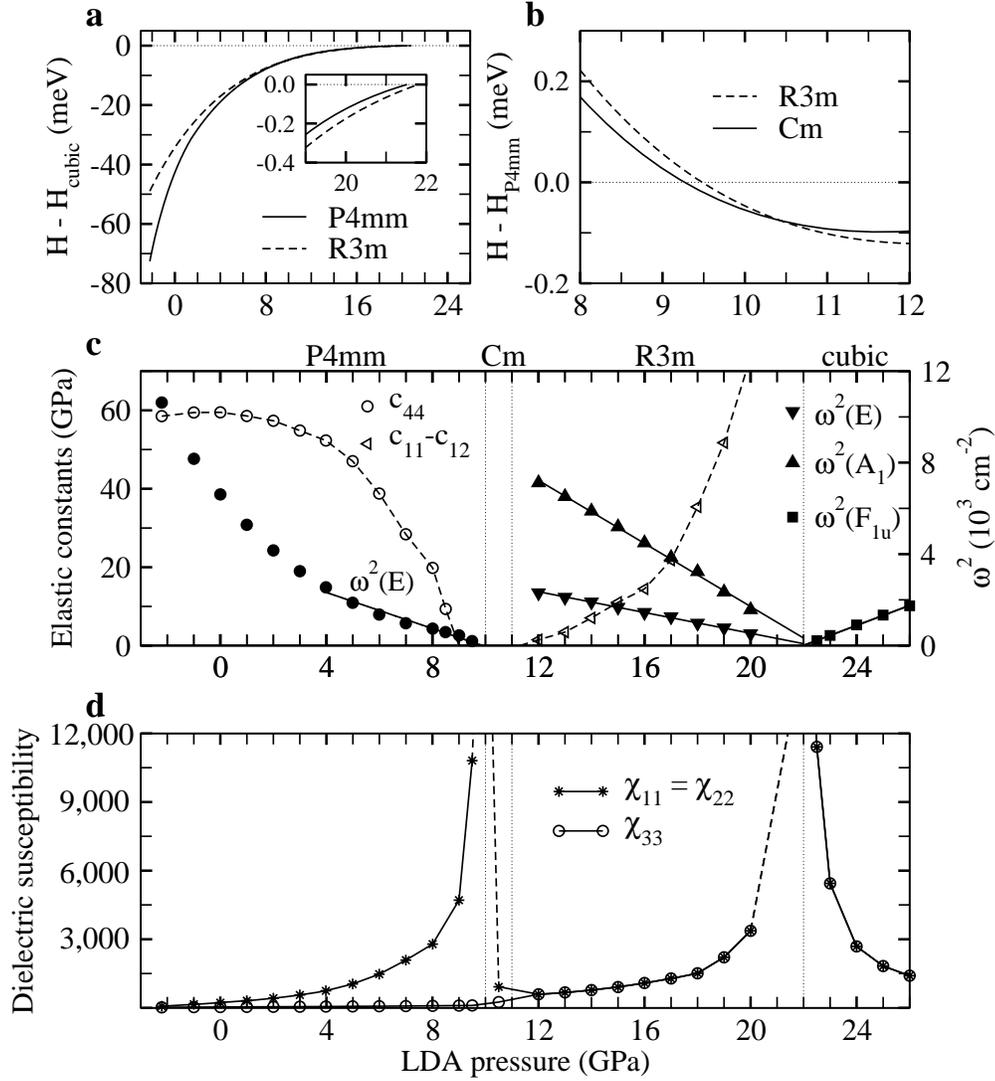

**Figure 5. Stability of PbTiO₃ phases and properties versus pressure. (a) Enthalpy difference with respect to the cubic (C) phase for the tetragonal (T) and rhombohedral (R) phases. (b) Enthalpy difference with respect to the T phase for the R and monoclinic (M) phases. (c) Elastic constants for the T and R phases, and square of the lowest optical phonon frequencies $\omega^2$ for the T, R, and C phases. (d) Dielectric susceptibility. From: [11].**

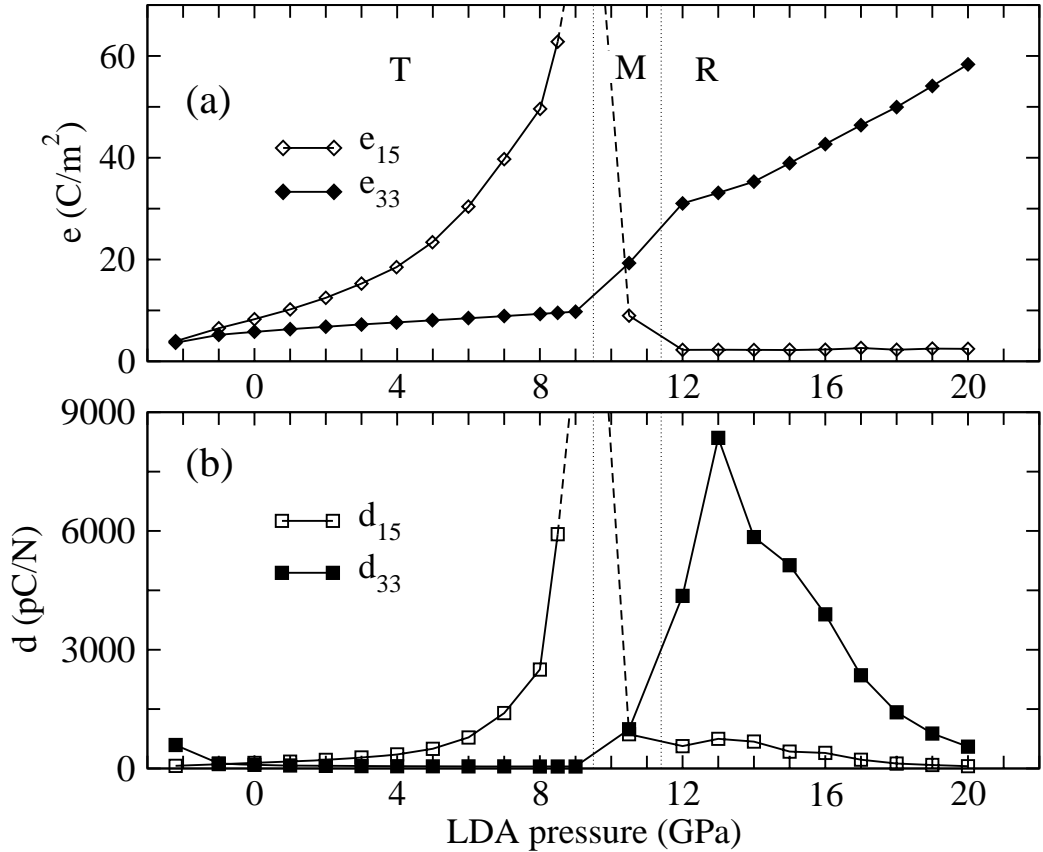

**Figure**
**stra**
**than**

A fe
bilit
an a
glas
dielectric response. Under field cooling, a ferroelectric phase transition is generally observed. As composition is varied and a ferroelectric component is added, a relaxor to ferroelectric phase transition occurs. Near the transition, the ferroelectrics retain some of the characteristics of relaxors such as dispersion or frequency dependence in properties. Relaxor ferroelectrics have been given much attention since the discovery of huge electromechanical coupling in single crystals such as $PbMg_{1/3}Nb_{2/3}O_3$ –$PbTiO_3$ (PMN-PT) and $PbZn_{1/3}Nb_{2/3}O_3$-$PbTiO_3$ (PZN-PT). [56] These crystals are rhombohedral at the relaxor end of the phase diagram, and tetragonal at the $PbTiO_3$ side, with a morphotropic phase boundary and intermediate monoclinic or orthorhombic phase in-between. The largest piezoelectric response occurs near the morphotropic phase boundary in the rhombohedral phase, with the polarization along the [111] directions, and an applied electric field along the [001] direction. The large response has been understood as due to polarization rotation. [17]

The polarization rotation effect can give much higher coupling than the normal collinear effect, in which the polarization and electric field are parallel to each other. The polarization is along [111] is compatible with a rhombohedral structure, along [001] with tetragonal, and along [011] with orthorhombic. Intermediate directions like [xx0] or [0x0] are monoclinic. When the polarization is rotated with an applied field, the response is like that of a field driven phase transition (Figs. 7-8). The large response in PMN-PT for example can be understood by considering pure PT. PT has a large tetragonal strain of 6%. If PT had a rhombohedral phase at zero pressure, one could apply a field to the rhombohedral phase driving the structure towards tetragonal. The net result would be a huge 6% strain. However, PT does not have a rhombohedral phase at zero pressure, but a solid solution with PMN, PZN, or PZ does have a rhombohedral phase that allows this rotation effect. The resulting strains are smaller than in pure PT, but still much larger than the collinear effect. The huge electromechanical coupling predicted in pure PT under pressure is due to polarization rotation as well.

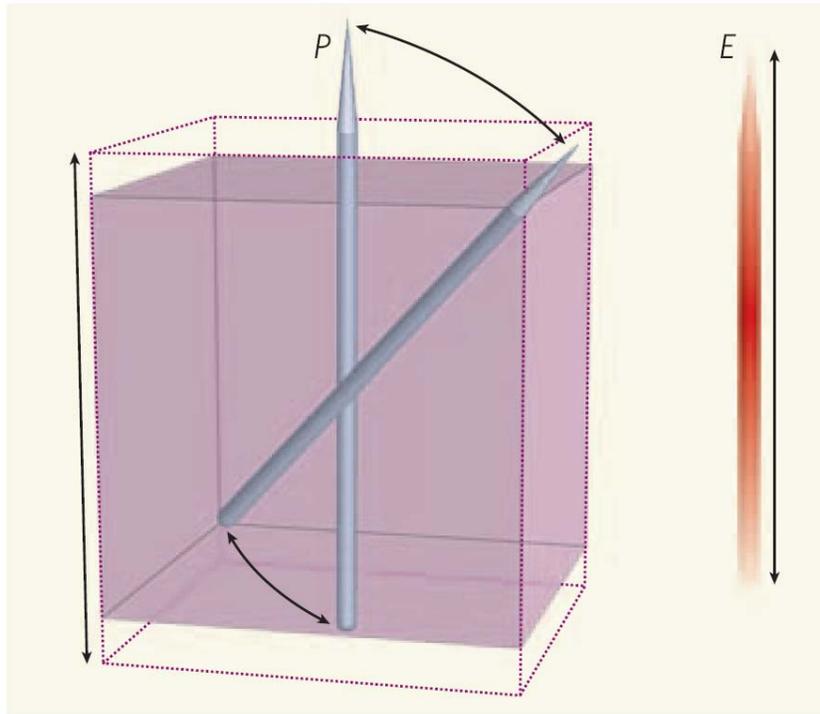

**Figure 7. Illustration of polarization rotation effect. An applied field along the cubic [001] direction rotations the polarization from [111], the rhombohedral direction, towards [001], through intermediate monoclinic states. From [57].**

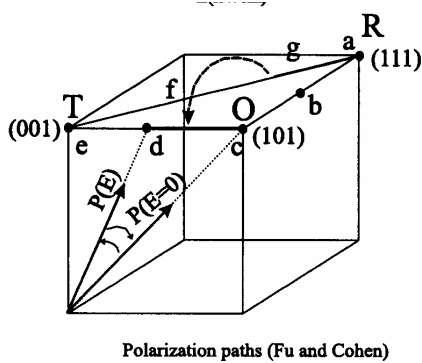

Polarization paths (Fu and Cohen)

**Figure 8. Representation of polarization directions. (a) rhombohedral symmetry [111] direction (b) monoclinic symmetric ($M_x$) [x01] (c) orthorhombic symmetric [101], (d) monoclinic symmetry [0x1] (e) tetragonal symmetry [001] (f) monoclinic symmetry [xx1] (g) triclinic [xy1].**

The polarization rotation effect has been verified now by numerous experiments. It leads to intermediate low symmetry monoclinic or orthorhombic phases between rhombohedral and tetragonal phases on the phase diagram. [58-63] Rotation has also been seen in situ measurements under applied electric fields. [64-66]

In addition to the polarization rotation, which gives rise to the large electromechanical coupling, relaxor ferroelectrics display a wealth of intriguing phenomena, especially in the relaxor endmember or close to it. These relaxors do not freeze into an ordered ground state, but instead display a broad frequency dependent maximum in the dielectric response. The generally accepted paradigm for these materials was outlined by Burns and Dacol, [67] who attributed non-linearity in the refractive index with temperature below a certain temperature as the onset of polarization in small nanoregions. Different polar nanoregions have polarizations pointing in different directions, and there is no long-range order among nanoregions which would give a phase transition. According to many views, these polar nanoregions are embedded in a non-polar matrix, but some data suggest that it might be better to regard the polarization correlation, which would show long tails rather than a clear separation between polar and non-polar regions. Theoretical models for disordered relaxors will be considered further below,

Relaxors can display two types of disorder. One type of disorder is the polarization disorder mentioned above. Another is chemical disorder. PMN is not generally found in a long-range ordered state, but there is evidence for small chemically ordered domains. [68-71] Other relaxors such as PST ($Pb(Sc_{0.5}Ta_{0.5})O_3$) can be chemically ordered by annealing, so that the effects of chemical order on their relaxor properties can be studied experimentally. In the ordered state PST has a normal ferroelectric phase transition, but shows relaxor behavior when chemically disordered. [72, 73] Statistical mechanical calculations show that the PST ground state is ordered; [74-76] the disordered state occurs from lack of equilibrium in the B-cation distribution caused by the slow diffusion times of the highly charged cations. The question remains if the chemical ordered state will always be ferroelectric in a relaxor, but it seems probable.

First-principles calculations for PMN do show a polar, ordered ground state with monoclinic C2 symmetry with 1:2 stacking of Mg and Nb along (111) planes [77] or 1:1 symmetry. [78, 79] If cubic PMN did form nanoscale monoclinic domains, it would be difficult to determine definitively using diffraction data since it is very difficult to deconvolute the diffraction data from a sample with 12 twin variants. Nanodomains of the twin variants would provide static random fields that would contribute further to the disorder of the material and possibly some of the frequency dependent relaxor behavior.

The effective charges in relaxors may contribute to the low symmetry ground state found for PMN. Computations of effective charges in PMN show them to be very anisotropic. Local fields in the polar phase will lead to oxygen displacements in oblique directions, giving rise to a monoclinic ground state.

### 3. LiNbO$_3$ and LiTaO$_3$

The lithium niobate structure is closely related to perovskite, and can be considered to be formed from perovskite by rotation of the octahedra, but the rotation is quite large the properties of ferroelectric lithium niobate and lithium tantalite are quite different from perovskite as a result of the different structure. The symmetry of the ferroelectric phase is R3c and thus they are uniaxial ferroelectrics. The soft mode is thus one-dimensional, and polarization rotation is not possible in this structure, greatly simplifying understanding of the ferroelectric phase transition. Complicating the picture is that Li is quite light, and thus can tunnel between the two wells in the double well potential surface. However, first-principles calculations show the wells to be quite deep (Fig. 9) and $T_c$ is high, so classical hopping dominates. Before the first-principles calculations were performed, fitting to experiments suggested a triple well potential surface, which is not consistent with the first-principles results. Frozen phonon calculations showed that the double well is largely due to hybridization between the $d_0$ Nb$^{5+}$ and Ta$^{5+}$ ions with oxygen, much as for the oxide perovskite, whereas previous work had considered the ferroelectricity to be mainly due to hopping of the Li. The Li does hop, but mainly in response to the changing fields from the off-centering of the Nb or Ta. Linear response computations gave the phone frequencies of the ground state ferroelectric state. [80] Modeling of the double well potential as an anharmonic oscillator gave good agreement with experiments. One outstanding problem is a first-principles understanding of the difference of Tc of LiNbO3 (1480 K) and LiTaO$_3$ (950 K) since they have almost identical potential surfaces.

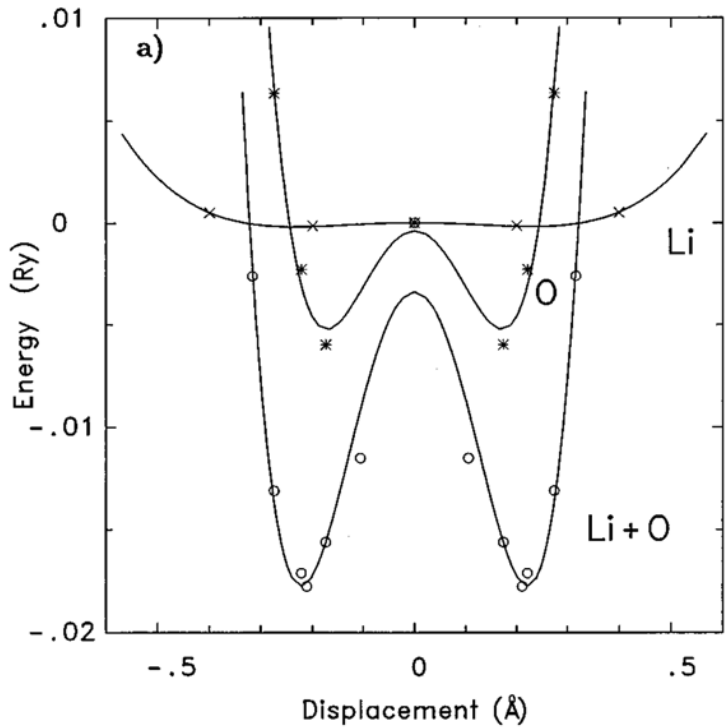

**Figure 9.** Computed energy surface for LiTaO$_3$ for displacing Li only, O only, or Li and O along the soft mode coordinate. Clearly the instability is not due to Li alone, and the concerted motion of Li and O is important. The results for LiNbO$_3$ are very similar. From [42].

## 4. GeTe and other IV-VI rocksalt chalcogenides

One of the first first-principles studies of a ferroelectric phase transition was the work of Rabe on GeTe. [7] This work gave the formation of an effective Hamiltonian and the statistical mechanics necessary to compute Tc and other thermal properties were developed. This landmark work has not yet been fully developed in more complex ferroelectrics, where instead efforts have focused on understanding of polarization and the use of molecular dynamics and Monte Carlo simulation techniques rather than renormalization group theory. Many of the important aspects of ferroelectrics were introduced in this study, including the importance of strain coupling, although the words "ferroelectric" or "piezoelectric" do not appear in the paper! GeTe remains a material with the simplest possible structure for a ferroelectric phase transition. There are only two atoms per unit cell, and the paraelectric phase is rocksalt structured. It would be worth revisiting such simple materials to better understand fundamental issues in ferroelectric and piezoelectric phenomena. More recently Waghmare et al. revisited GeTe and other chalcogenides, using first-principles techniques developed in the last 20 years that were unavailable in the 1980's, including linear response and Barry's phase techniques [81]. They found large effective charges and large optical dielectric constants, and showed that the ferroelectric instabilities are due partly to lone pair activity.

## 5. KDP

KH$_2$PO$_4$ (KPD) is a complex so-called "hydrogen-bonded" ferroelectric. Although it is very hard to understand how it works, KDP can easily be grown in large crystals suitable for non-linear optic applications. There is a large theoretical and experimental literature which is impossi-ble to review here, but very briefly, there are tetrahedral phosphate groups that are bonded to each other with hydrogen bonds. The groups can librate and the hydrogen bonds obey the Slater ice rules [82, 83]. The hydrogen orientations order below a $T_c$ of 122 K, all being bound to the top (+c) or bottom (-c) tetrahedral edges, giving rise to a net polarization. There is a huge negative isotope effect, with the deuterated form of KDP (DKDP) having a $T_c$ of 229 K. Deuteration changes the proton (deuteron) dynamics, possibly changing tunneling probabilities, and changes the crystal structural parameters. Koval et al. performed a detailed DFT study of KDP and shed some light on the nature of the ferroelectric phase transition in KDP and many long standing issues are discussed, including the origin of the polarization, the relative importance of proton motions and the heavy ions, and the importance of tunneling. [84] Contrary to many studies they concluded that tunneling is not of major importance in KDP and that the large isotope effect is due primarily to changes in the structural geometry on deuteration, due to changes in the zero point motion of the protons (deuterons) but not due to tunneling among the different wells. KDP is a good example of how first-principles methods can help clear up long standing mysteries that remained elusive using only experimental data and phenomenological models.

## 6. Multiferroics

Multiferroics are magnetic polar materials, which have cross coupling between magnetism and polarization [85, 86]. Usually magnetism and ferroelectricity are incompatible, because ferroelectricity is usually driven by hybridization between O 2p states and d-states that are the lowest unoccupied states, giving a $d_0$ configuration, whereas magnetism depends on partially occupied d (or f) states. One material which has received much attention as a magnetic ferroelectric is $YMnO_3$ [87, 88], which is an improper ferroelectric antiferromagnet. This material is a theoretical challenge because conventional DFT gives a metallic band structure, and it is necessary to use self-interaction corrections (SIC) [87], LDA+U [88] or other technique which includes local Coulomb interactions to obtain the proper insulating state. $YMnO_3$ is probably not a route to a useful multiferroic, as it has a small polarization of only 6.2 $\mu C/cm^2$ and is an antiferromagnetic, yet it points out several interesting materials properties. It is an improper ferroelectric, in that it is a zone boundary instability that is the primary order parameter. It is also a good case example of how theory can be useful in explaining ambiguous experimental data. Contrary to most primary ferroelectrics, $YMnO_3$ does not have enhanced transverse effective charges, and there is little or no displacement of the Mn relative to its neighboring oxygens. The small polarization is primarily due to displacements of the Y ions in response to octahedral tilts.

Theory is on the forefront of the multiferroic field, and one candidate material, $BiFeO_3$, has received much experimental and theoretical attention. [89-91]. Theory has been indispensable in understanding experimental data. Particularly important was the theoretical determination of the magnitude of polarization in $BiFeO_3$, which has been discrepant between thin film and bulk measurements. Another important and crucial theoretical contribution was an understanding of how magnetization direction could be switched by switching the polarization direction, a result that had incorrectly appeared to be unallowed by symmetry. Experiments continue on bismuth ferrite [92] and theory will be needed to explain, for example, the phonon anomalies observed at the Néel temperature. The observed complete absence of a Raman active soft mode at the transition should also be examined theoretically. Singh [93] studied the multiferroic $PbVO_3$ and showed the origin of its insulating state, magnetism, and polar ground state. Very recently Fennie and Rabe designed a new multiferroic, $EuTiO_3$, using first-principles theory, [94], bringing us to the final topic of "Materials by Design."

## 7. Materials by Design

One of the key goals of first-principles theory on piezoelectrics is to develop understanding and methods for carrying out computational materials by design. Some day it will be possible to de-sign materials of desired properties computationally, which then can be produced in the laboratory for testing, conformation, and development.

One example of such work is that of Singh work on new materials. Singh et al. [95] used first-principles methods applied to a variety of oxide perovskite ferroelectrics to come up with a design plan for new materials. Such an approach shows a broader use of first-principles methods to gain deep understanding for materials design, as opposed to predicting properties only for certain candidate materials. Multiferroics also provide an important example of materials by design. The prediction of $Bi_2FeCrO_6$ as a large polarization multiferroic [96] is very exciting, and an example of modern materials by design at its best. Other examples have been given above. The prediction of ferroelectric superlattices and their realization is another major advance in the "Materials by Design" concept. [28-31]

### *C. Summary*

First-principles theory has contributed significantly to our understanding of piezoelectrics and ferroelectrics. This is a rapidly growing field. We can expect theoretical methods to continue to get more robust and accurate, and to address more complex properties of piezoelectric materials. Key advances that came from theory are (1) understanding the role of hybridization and covalency in ferroelectric instability, (2) the role of large effective charges in electromechanical coupling, (3) a fundamental understanding of macroscopic polarization, (4) the role of polarization rotation in the single crystal relaxor ferroelectrics, (5) the concept of polar ferroelectric superlatcies, (6) an understanding of the requirements for materials to be multiferroics, and (7) design paths for new materials.

### *References*

1. Devonshire, A.F., *Theory of Ferroelectrics.* Philosophical Magazine, 1954. **Advances in Physics 3**(10).
2. Vanderbilt, D. and M.H. Cohen, *Monoclinic and triclinic phases in higher-order Devonshire theory.* Phys. Rev. B, 2001. **63**: p. 94108-94117.
3. Boyer, L.L., et al., *First principles calculations for ferroelectrics -- A vision.* Ferroelec., 1990. **111**: p. 1-7.
4. Cohen, R.E. and H. Krakauer, *Lattice dynamics and origin of ferroelectricity in BaTiO3: Linearized augmented plane wave total energy calculations.* Phys. Rev. B, 1990. **42**(10): p. 6416-6423.
5. Cohen, R.E., *Origin of ferroelectricity in oxide ferroelectrics.* Nature, 1992. **358**: p. 136-138.
6. Cohen, R.E., *Ferroelectricity origins.* Nature, 1993. **362**: p. 213.
7. Rabe, K.M. and J.D. Joannopoulos, *Theory of the structural phase transition of GeTe.* Phys. Rev. B, 1987. **36**(12): p. 6631-6639.
8. Saghi-Szabo, G., R.E. Cohen, and H. Krakauer, *First-principles study of piezoelectricity in PbTiO3.* Phys. Rev. Lett., 1998. **80**(19): p. 4321-4324.
9. Saghi-Szabo, G., R.E. Cohen, and H. Krakauer, *First-principles study of piezoelectricity in tetragonal PbTiO3 and PbZr1/2Ti1/2O3.* Phys. Rev. B, 1999. **59**: p. 12771-12776.
10. Wu, Z., et al., *Erratum: First-Principles Study of Piezoelectricity in PbTiO[sub 3] [Phys. Rev. Lett. [bold 80], 004321 (1998)].* Phys. Rev. Lett., 2005. **94**(6): p. 069901.
11. Wu, Z. and R.E. Cohen, *Pressure-Induced Anomalous Phase Transitions and Colossal Enhancement of Piezoelectricity in PbTiO3.* Phys. Rev. Lett., 2005. **95**: p. 037601.
12. Resta, R., *Ab initio simulation of the properties of ferroelectric materials.* Model. Simul. Mat. Sci. Eng., 2003. **11**(4): p. R69-R96.

13. Rabe, K.M. and P. Ghosez, *First-principles studies of ferroelectric oxides*, in *Physics of Ferroelectrics: a Modern Perspective*, C.H. Ahn, K.M. Rabe, and J.M. Triscone, Editors. 2006, Springer: New York.
14. Fiolhais, C., F. Nogueira, and M. Marques, eds. *A Primer in Density Functional Theory*. Lecture Notes in Physics. Vol. 620. 2003, Springer: New York. 256.
15. Kohn, W. and L.J. Sham, *Self-consistent equations including exchange and correlation effects.* Phys. Rev. A, 1965. **140**: p. 1133-1140.
16. Slater, J.C., *The Lorenz correction in Barium Titanate.* Phys. Rev., 1950. **78**(6): p. 748-761.
17. Fu, H. and R.E. Cohen, *Polarization rotation mechanism for ultrahigh electromechanical response in single-crystal piezoelectrics.* Nature, 2000. **403**: p. 281-283.
18. Burton, B.P., E. Cockayne, and U.V. Waghmare, *Correlations between nanoscale chemical and polar order in relaxor ferroelectrics and the lengthscale for polar nanoregions.* Phys. Rev. B, 2005. **72**(6): p. 064113.
19. Tinte, S., et al., *Origin of the Relaxor State in Pb(B[sub x]B[sub 1-x][sup [prime]])O[sub 3] Perovskites.* Phys. Rev. Lett., 2006. **97**(13): p. 137601-4.
20. Kornev, I.A., et al., *Ferroelectricity of Perovskites under Pressure.* Phys. Rev. Lett., 2005. **95**(19): p. 196804.
21. King-Smith, R.D. and D. Vanderbilt, *Theory of polarization of crystalline solids.* Phys. Rev. B, 1993. **47**(3): p. 1651-1654.
22. Vanderbilt, D. and R.D. King-Smith, *Electric polarization as a bulk quantity and its relation to surface charge.* Phys. Rev. B, 1993. **48**(7): p. 4442-4455.
23. Resta, R., *Macroscopic polarization in crystalline dielectrics: the geometric phase approach.* Rev. Mod. Phys., 1994. **66**: p. 899-915.
24. Resta, R., *Polarization Fluctuations in Insulators and Metals: New and Old Theories Merge.* Phys. Rev. Lett., 2006. **96**(13): p. 137601.
25. Resta, R., *Why are insulators insulating and metals conducting?* J Phys. Cond. Matt., 2002. **14**(20): p. R625-R656.
26. Souza, I., J. Iniguez, and D. Vanderbilt, *First-Principles approach to insulators in finite electric fields.* Phys. Rev. Lett., 2002. **89**: p. 117602.
27. Souza, I., J. Iniguez, and D. Vanderbilt, *Dynamics of Berry-phase polarization in time-dependent electric fields.* Phys. Rev. B, 2004. **69**(8): p. 851061.
28. Bungaro, C. and K.M. Rabe, *Lattice instabilities of PbZrO3/PbTiO3 [1:1] superlattices from first principles.* Phys. Rev. B, 2002. **6522**(22): p. 4106-4106.
29. Warusawithana, M.P., et al., *Artificial dielectric superlattices with broken inversion symmetry.* Phys. Rev. Lett., 2003. **90**: p. 036802.
30. Dawber, M., et al., *Unusual behavior of the ferroelectric polarization in PbTiO3/SrTiO3 superlattices.* Phys. Rev. Lett., 2005. **95**(17): p. 1-4.
31. Nakhmanson, S.M., K.M. Rabe, and D. Vanderbilt, *Predicting polarization enhancement in multicomponent ferroelectric superlattices.* Phys. Rev. B, 2006. **73**(6): p. 1-4.
32. Wu, Z. and R.E. Cohen, *More accurate generalized gradient approximation for solids.* Phys. Rev. B, 2006. **73**(23): p. 235116-6.
33. Gonze, X., D.C. Allan, and M.P. Teter, *Dielectric tensor, effective charges, and phonons in à-quartz by variational density-functional perturbation theory.* Phys. Rev. Lett., 1992. **68**(24): p. 3603-3606.
34. Gonze, X. and C. Lee, *Dynamical matrices, Born effective charges, dielectric permittivity tensors, and interatomic force constants from density-functional perturbation theory.* Phys. Rev. B, 1997. **55**(16): p. 10355–10368.
35. Hamann, D.R., et al., *Metric tensor formulation of strain in density-functional perturbation theory.* Phys. Rev. B, 2005. **71**(3): p. 035117.

36. Ghosez, P., J.P. Michenaud, and X. Gonze, *Dynamical atomic charges: The case of ABO3 compounds.* Phys. Rev. B, 1998. **58**(10): p. 6224-6240.
37. Axe, J.D., *Apparent ionic charges and vibrational eigenmodes of BaTiO3 and other perovskites.* Phys. Rev., 1967. **157**: p. 429-435.
38. Scott, J.F., *Raman Spectra and Lattice Dynamics of alpha-Berlinite (AlPO[sub 4]).* Phys. Rev. B, 1971. **4**(4): p. 1360-1366.
39. Zhong, W., D. Vanderbilt, and K.M. Rabe, *Phase transitions in BaTiO3 from first principles.* Phys. Rev. Lett., 1994. **73**(13): p. 1861-1864.
40. Rabe, K.M. and U.V. Waghmare, *Localized basis for effective lattice Hamiltonians: Lattice Wannier functions.* Phys. Rev. B, 1996. **52**: p. 13236-13246.
41. Waghmare, U.V. and K.M. Rabe, *Ab initio statistical mechanics of the ferroelectric phase transition in PbTiO3.* Phys. Rev. B, 1997. **55**: p. 6161-6173.
42. Inbar, I. and R.E. Cohen, *Comparison of the electronic structures and energetics of LiTaO3 and LiNbO3.* Phys. Rev. B, 1994. **53**: p. 1193-1204.
43. Luspin, Y., J.L. Servoin, and F. Gervais, *Soft mode spectroscopy in barium titanate.* J. Phys. C: Solid St. Phys., 1980. **13**: p. 3761-73.
44. Comes, R., M. Lambert, and A. Guinier, *The chain structure of BaTiO3 and KNbO3.* Solid State Comm., 1968. **6**: p. 715-719.
45. Sicron, N., et al., *Nature of the ferroelectric phase transition in PbTiO$_{3}$.* Phys. Rev. B, 1994. **50**(18): p. 13168.
46. Rossetti Jr, G.A., J.P. Cline, and A. Navrotsky, *Phase transition energetics and thermodynamic properties of ferroelectric PbTiO3.* J. Mat. Res., 1998. **13**(11): p. 3197-3206.
47. Ghosez, P., et al., *Lattice dynamics of BaTiO3, PbTiO3, and PbZrO3: A comparative first-principles study.* Phys. Rev. B, 1999. **60**: p. 836-843.
48. Singh, D.J., *Structure and energetics of Antiferroelectric PbZrO3.* Phys. Rev. B, 1995. **52**: p. 12559-12563.
49. Fornari, M. and D.J. Singh, *Possible coexistence of rotational and ferroelectric lattice distortions in rhombohedral PbZrxTi1-xO3.* Phys. Rev. B, 2001. **63**(9): p. 921011.
50. Ghita, M., et al., *Interplay between A-site and B-site driven instabilities in perovskites.* Phys. Rev. B, 2005. **72**(5).
51. Johannes, M.D. and D.J. Singh, *Crystal structure and electric field gradients of PbZrO3 from density functional calculations.* Phys. Rev. B, 2005. **71**(21): p. 1-4.
52. Resta, R. and D. Vanderbilt, *Theory of Polarization: A Modern Approach*, in *Physics of Ferroelectrics: a Modern Perspective*, C.H. Ahn, K.M. Rabe, and J.M. Triscone, Editors. 2007, Springer-Verlag: New York.
53. Boyer, L., H. Stokes, and M. Mehl, *Calculation of polarization using a density functional method with localized charge.* Phys. Rev. Lett., 2000. **84**(4): p. 709-712.
54. Choudhury, N., R.E. Cohen, and E.J. Walter, *First principles studies of the Born effective charges and electronic dielectric tensors for the relaxor PMN (PbMg1/3Nb2/3O3).* Computational Materials Science, 2006. **37**: p. 152.
55. Posternak, M., R. Resta, and A. Baldereschi, *Role of covalent bonding in the polarization of perovskite oxides: the case of KNbO3.* Phys. Rev. B, 1994. **50**: p. 8911-8914.
56. Park, S.E. and T.R. Shrout, *Ultrahigh strain and piezoelectric behavior in relaxor based ferroelectric single crystals.* J. Appl. Phys., 1997. **82**(4): p. 1804-1811.
57. Cohen, R.E., *Relaxors go critical.* Nature, 2006. **441**: p. 941.
58. La-Orauttapong, D., et al., *Phase diagram of the relaxor ferroelectric (1-x)Pb(Zn1/3Nb2/3)O-3-xPbTiO(3).* Phys. Rev. B, 2002. **65**(14): p. 4101-4101.
59. Ohwada, K., et al., *Neutron diffraction study of the irreversible R-M$_A$-M$_C$ phase transition in single crystal Pb[(Zn$_{1/3}$Nb$_{2/3}$)$_{1-}$* $x$Ti$_x$]O$_3$. Journal of the Physical Society of Japan, 2001. **70**(9): p. 2778.
60. Noheda, B., et al., *Tetragonal-to-monoclinic phase transition in a ferroelectric perovskite: The structure of PbZr0.52Ti0.48O3.* Phys. Rev. B, 2000. **61**(13): p. 8687-8695.
61. Noheda, B., et al., *New features of the morphotropic phase boundary in the Pb(Zr1-xTix)O-3 system.* Ferroelec., 2000. **237**(1-4): p. 541-548.
62. Guo, R., et al., *Origin of the high piezoelectric response in PbZr1-xTixO3.* Phys. Rev. Lett., 2000. **84**(23): p. 5423-5426.
63. Noheda, B., et al., *A monoclinic ferroelectric phase in the Pb(Zr1-xTix)O-3 solid solution.* Appl. Phys. Lett., 1999. **74**(14): p. 2059-2061.
64. Ohwada, K., et al., *Neutron diffraction study of field-cooling effects on the relaxor ferroelectric Pb[(Zn1/3Nb2/3)(0.92)Ti-0.08]O-3 - art. no. 094111.* Phys. Rev. B, 2003. **6709**(9): p. 4111-4111.
65. Noheda, B., et al., *Electric-field-induced phase transitions in rhombohedral Pb(Zn1/3Nb2/3)(1-x)TixO3 - art. no. 224101.* Phys. Rev. B, 2002. **6522**(22): p. 4101-4101.
66. Noheda, B., et al., *Polarization rotation via a monoclinic phase in the piezoelectric 92% PbZn1/3Nb2/3O3-8% PbTiO3.* Phys. Rev. Lett., 2001. **86**(17): p. 3891-3894.
67. Burns, G. and F.H. Dacol, *Crystalline ferroelectrics with glassy polarization behavior.* Phys. Rev. B, 1983. **28**(5): p. 2527.
68. Wakimoto, S., et al., *Ferroelectric ordering in the relaxor Pb(Mg1/3Nb2/3)O-3 as evidenced by low-temperature phonon anomalies* Phys. Rev. B, 2002. **65**: p. 172105.
69. Davies, P.K., et al., *Cation ordering and dielectric properties of PMN-PNS relaxors.* 2000.
70. Davies, P.K. and M.A. Akbas, *Chemical order in PMN-related relaxors: Structure, stability, modification, and impact on properties.* J. Phys. Chem. Solids, 2000. **61**(2): p. 159.
71. Egami, T., et al., *Nature of atomic ordering and mechanism of relaxor ferroelectric phenomena in PMN.* Ferroelec., 1998. **206-207**(1 -4; 1-2): p. 231.
72. Farber, L., et al., *Cation ordering in Pb(Mg1/3Nb2/3)O3-Pb(Sc1/2Nb1/2 )O3 (PMN-PSN) solid solutions.* J. Am. Ceram. Soc., 2002. **85**(9): p. 2319.
73. Farber, L. and P. Davies, *Influence of Cation Order on the Dielectric Properties of Pb(Mg 1/3Nb2/3)O3-Pb(Sc1/2Nb 1/2)O3 (PMN-PSN) Relaxor Ferroelectrics.* J. Am. Ceram. Soc., 2003. **86**(11): p. 1861.
74. Burton, B.P. and R.E. Cohen, *Theoretical study of cation ordering in the system Pb(Sc$_{1/2}$Ta$_{1/2}$)O$_3$.* Ferroelec., 1994. **151**: p. 331-336.
75. Burton, B.P. and R.E. Cohen, *Non-empirical calculation of the Pb(Sc0.5Ta0.5)O3-PbTiO3 quasibinary phase diagram.* Phys. Rev. B, 1995. **52**(2): p. 792-797.
76. Burton, B.P. and R.E. Cohen, *First-principles study of cation ordering in the system PbSc1/2Ta1/2)O3 and (1-X) Pb(Sc1/2Ta1/2)O3-X PbTiO3.* Ferroelec., 1995. **164**: p. 201-212.
77. Choudhury, N., et al., *Ab initio linear response and frozen phonons for the relaxor PbMg$_{1/3}$Nb$_{2/3}$O$_3$.* Phys. Rev. B, 2005. **71**(12): p. 125134.
78. Prosandeev, S.A., et al., *Lattice dynamics in PbMg$_{1/3}$Nb$_{2/3}$O$_3$.* Phys. Rev. B, 2004. **70**(13): p. 134110.
79. Suewattana, M. and D.J. Singh, *Electronic structure and lattice distortions in PbMg$_{1/3}$Nb$_{2/3}$O$_3$ studied with density functional theory using the linearized augmented plane-wave method.* Phys. Rev. B, 2006. **73**(22): p. 224105-5.
80. Veithen, M. and P. Ghosez, *First-principles study of the dielectric and dynamical properties of lithium niobate.* Phys. Rev. B, 2002. **65**(21): p. 214302.

81. Waghmare, U.V., et al., *First-principles indicators of metallicity and cation off-centricity in the IV-VI rocksalt chalcogenides of divalent Ge, Sn, and Pb.* Phys. Rev. B, 2003. **67**(12): p. 125111-10.
82. Slater, J.C., *Theory of the Transition in KH[sub 2]PO[sub 4].* The Journal of Chemical Physics, 1941. **9**(1): p. 16-33.
83. Kind, R., et al., *Slater Ice Rules and H-Bond Dynamics in KDP-Type Systems.* Phys. Rev. Lett., 2002. **88**(19): p. 195501.
84. Koval, S., et al., *First-principles study of ferroelectricity and isotope effects in H-bonded K H2 P O4 crystals.* Phys. Rev. B, 2005. **71**(18): p. 1-15.
85. Hill, N.A., *Density functional studies of multiferroic magnetoelectrics.* Ann. Rev. Mater. Res., 2002. **32**: p. 1-37.
86. Ederer, C. and N.A. Spaldin, *Recent progress in first-principles studies of magnetoelectric multiferroics.* Current Opinion in Solid State and Materials Science, 2005. **9**(3): p. 128-139.
87. Van Aken, B.B., et al., *The origin of ferroelectricity in magnetoelectric YMnO3.* Nature Materials, 2004. **3**(3): p. 164-170.
88. Fennie, C.J. and K.M. Rabe, *Ferroelectric transition in YMnO[sub 3] from first principles.* Phys. Rev. B, 2005. **72**(10): p. 100103.
89. Neaton, J.B., et al., *First-principles study of spontaneous polarization in multiferroic BiFeO[sub 3].* Phys. Rev. B, 2005. **71**(1): p. 014113.
90. Ederer, C. and N.A. Spaldin, *Weak ferromagnetism and magnetoelectric coupling in bismuth ferrite.* Phys. Rev. B, 2005. **71**(6): p. 060401.
91. Baettig, P., C. Ederer, and N.A. Spaldin, *First principles study of the multiferroics BiFeO3, Bi2FeCrO6, and BiCrO3: Structure, polarization, and magnetic ordering temperature.* Phys. Rev. B, 2005. **72**(21): p. 1-8.
92. Haumont, R., et al., *Phonon anomalies and the ferroelectric phase transition in multiferroic BiFeO[sub 3].* Phys. Rev. B, 2006. **73**(13): p. 132101-4.
93. Singh, D.J., *Electronic structure and bond competition in the polar magnet PbVO[sub 3].* Phys. Rev. B, 2006. **73**(9): p. 094102-5.
94. Fennie, C.J. and K.M. Rabe, *Magnetic and electric phase control in epitaxial EuTiO$_3$ from first principles.* Phys. Rev. Lett., 2006. **in press**.
95. Singh, D.J., et al., *The role of Pb in piezoelectrics and possible substitutions for it.* Journal De Physique. IV, 2005. **128**: p. 47-53.
96. Baettig, P. and N.A. Spaldin, *Ab initio prediction of a multiferroic with large polarization and magnetization.* Appl. Phys. Lett., 2005. **86**(1): p. 012505-3.